\documentclass[aps,prb,a4paper,superscriptaddress,twocolumn,showpacs,amsmath,amssymb]{revtex4} 
\usepackage{graphicx,epsfig} 
\allowdisplaybreaks
\begin{document} 

\newcommand{\s}{\sigma}
\newcommand{\up}{\uparrow}
\newcommand{\dw}{\downarrow}
\newcommand{\h}{\mathcal{H}}
\newcommand{\g}{\mathcal{G}^{-1}_0}
\newcommand{\D}{\mathcal{D}}
\newcommand{\A}{\mathcal{A}}
\newcommand{\K}{\textbf{k}}
\newcommand{\Q}{\textbf{q}}
\newcommand{\T}{\tau_{\ast}}
\newcommand{\io}{i\omega_n}
\newcommand{\eps}{\varepsilon}
\newcommand{\+}{\dag}
\newcommand{\su}{\uparrow}
\newcommand{\giu}{\downarrow}
\newcommand{\0}[1]{\textbf{#1}}
\newcommand{\ca}{c^{\phantom{\dagger}}}
\newcommand{\cc}{c^\dagger}
\newcommand{\da}{d^{\phantom{\dagger}}}
\newcommand{\dc}{d^\dagger}
\newcommand{\be}{\begin{equation}}
\newcommand{\ee}{\end{equation}}
\newcommand{\bea}{\begin{eqnarray}}
\newcommand{\eea}{\end{eqnarray}}
\newcommand{\ba}{\begin{eqnarray*}}
\newcommand{\ea}{\end{eqnarray*}}
\newcommand{\dagga}{{\phantom{\dagger}}}
\newcommand{\bR}{\mathbf{R}}
\newcommand{\bQ}{\mathbf{Q}}
\newcommand{\bq}{\mathbf{q}}
\newcommand{\bqp}{\mathbf{q'}}
\newcommand{\bk}{\mathbf{k}}
\newcommand{\bh}{\mathbf{h}}
\newcommand{\bkp}{\mathbf{k'}}
\newcommand{\bp}{\mathbf{p}}
\newcommand{\bRp}{\mathbf{R'}}
\newcommand{\bx}{\mathbf{x}}
\newcommand{\by}{\mathbf{y}}
\newcommand{\bz}{\mathbf{z}}
\newcommand{\br}{\mathbf{r}}
\newcommand{\Ima}{{\Im m}}
\newcommand{\Rea}{{\Re e}}
\newcommand{\Pj}[2]{|#1\rangle\langle #2|}
\newcommand{\ket}[1]{|#1\rangle}
\newcommand{\fract}[2]{\frac{\displaystyle #1}{\displaystyle #2}}
\newcommand{\Av}[2]{\langle #1|\,#2\,|#1\rangle}
\newcommand{\eqn}[1]{(\ref{#1})}
\newcommand{\Tr}{\mathrm{Tr}}
\title{Fermi-surface evolution across the magnetic phase transition in the Kondo lattice model}
\author{Nicola Lanat\`a}
\affiliation{International School for Advanced Studies (SISSA), and CRS Democritos, CNR-INFM,
Via Beirut 2-4, I-34014 Trieste, Italy} 
\author{Paolo Barone}
\affiliation{International School for Advanced Studies (SISSA), and CRS Democritos, CNR-INFM,
Via Beirut 2-4, I-34014 Trieste, Italy} 
\author{Michele Fabrizio} 
\affiliation{International School for Advanced Studies (SISSA), and CRS Democritos, CNR-INFM,
Via Beirut 2-4, I-34014 Trieste, Italy}
\affiliation{The Abdus Salam International Centre for Theoretical Physics 
(ICTP), P.O.Box 586, I-34014 Trieste, Italy} 
\date{\today} 
\begin{abstract}
We derive, by means of an extended Gutzwiller wavefunction and within the Gutzwiller approximation, the phase diagram 
of the Kondo lattice model. We find that generically, namely in the absence of nesting, the model displays an $f$-electron  
Mott localization accompanied by a discontinuous change of the conduction electron Fermi surface as well as by magnetism. 
When the non interacting Fermi surface is close to nesting, the Mott localization disentangles from the onset of magnetism. 
First the paramagnetic heavy fermion metal turns continuously into an itinerant magnet - the Fermi surface evolves 
smoothly across the transition - and afterwards Mott localization intervenes with a discontinuous rearrangement of the Fermi 
surface. We find that the $f$-electron localization remains even if magnetism is prevented, and is still accompanied by a 
sharp transfer of spectral weigth at the Fermi energy within the Brillouin zone. 
We further show that the Mott localization can be also induced by an external 
magnetic field, in which case it occurs concomitantly with a metamagnetic transition.     
\end{abstract}

\maketitle

\section{Introduction}

The physics of heavy-fermion compounds remains a fascinating and challenging issue within strongly correlated 
materials. Recently, considerable experimental and theoretical interest has focused on the physical behavior across the 
magnetic quantum phase transition that is traditionally expected to occur when the Ruderman-Kittel-Kasuya-Yosida (RKKY) interaction overwhelms Kondo 
screening.~\cite{Doniach} This transition is induced experimentally by external parameters like chemical composition, 
pressure or magnetic field, see for instance Refs.~\onlinecite{Stewart} and \onlinecite{Vojta} as well as references therein, 
and is commonly accompanied by topological changes of the Fermi 
surface~\cite{Aoki,nature-si,Canfield,Onuki,Julian} and anomalous behavior of various transport and thermodynamic 
quantities.~\cite{Stewart,Vojta} The theoretical debate on this subject 
has so far mainly followed two 
different directions.~\cite{Vojta} One ascribes the changes of the Fermi surface to 
an $f$-electron Mott localization,~\cite{Johansson} which is assumed to occur concomitantly with magnetism as well as with the 
appearance of transport and thermodynamics anomalies.~\cite{Coleman-Pepin,Si,Pepin,Pepin-2} The alternative proposal 
assumes that magnetism is predominantly an instability of an itinerant phase~\cite{Allen,Lavagna}, hence that the 
Fermi surface changes arise simply by the spin polarization of dispersing bands~\cite{Allen,Lavagna} and the anomalous behavior 
by critical magnetic quantum fluctuations.~\cite{millis-1993}
This issue has been very recently addressed theoretically in the periodic Anderson model  
by De Leo, Civelli and Kotliar~\cite{DeLeo,deleo-2008} using a cluster extension of dynamical mean field theory (CDMFT). 
Upon decreasing the hybridization between $f$-orbitals and conduction electrons, a weak first 
order phase transition from a heavy-fermion paramagnet to an itinerant antiferromagnet has been found. Remarkably, when these authors 
force CDMFT not to break spin SU(2) symmetry and follow the metastable paramagnetic solution, they find an 
orbital-selective Mott localization - a pseudogap opens in the $f$-electron spectral function 
at the chemical potential, although low energy spectral weight remains within the Mott-Hubbard gap~\cite{FerreroPRB,Georges-OSMT} - 
for a hybridization between $f$ and conduction electrons almost coincident with the value at which, allowing for magnetism, 
the antiferromagnetic transition occurs.
This result suggests that the magnetic phase transition masks an incipient 
Mott localization of the $f$-electrons, which could become visible above the Ne\`el temperature or by suppressing antiferromagnetism. 
A complementary attempt has been almost contemporarily performed by Watanabe and Ogata.~\cite{Ogata} 
These authors analyse by a variational Monte-Carlo (VMC) technique a Gutzwiller wavefunction for a Kondo lattice model in 
a two-dimensional square lattice. The variational phase diagram as function of the Kondo exchange depends non-trivially 
on the electron density. Very close to the compensated regime (one conduction electron per impurity-spin), 
upon decreasing the Kondo exchange there is first a second-order paramagnetic-to-antiferromagnetic phase transition, followed 
by a first-order transition between two magnetic phases with different Fermi surfaces. Moving away from 
the compensated regime, the second order phase transition disappears and they find a single first order line separating  
a paramagnetic phase from an antiferromagnetic one with different Fermi surfaces. These VMC results suggest that the magnetic 
transition and the topological change of the Fermi surface are not necessarily coincident, which has been also observed 
in a very recent experiment.~\cite{Paglione} Since a variational calculation can only access ground state properties and 
not subtle dynamical features like an orbital-selective Mott transition, and keeping into account the differences between 
the periodic Anderson model and the Kondo lattice model, the VMC~\cite{Ogata} and CDMFT~\cite{DeLeo} 
results might not be incompatible one to the other, and instead describe the same physical scenario although from 
two different perspectives. Should this be the case, it would undoubtly represent  
a step forward in the comprehension of heavy-fermion physics. To settle this question, 
one should for instance try to get closer to the compensated regime by CDMFT and check whether 
the $f$-localization and the on-set of magnetism disentangle from each other as predicted by VMC. Alternatively, one could 
carry on with variational calculations trying to uncover features that indirectly signal the $f$-localization. This is the aim of  
the present work. We note, by the way, that finite average values of the hybridization between $f$ orbitals and conduction electrons, 
in the periodic Anderson model, or of the Kondo exchange, in the Kondo lattice model, must not be interpreted as 
absence of $f$-localization in a proper variational calculation, since the hybridization or the Kondo exchange 
are part of the Hamiltonian. Therefore other quantities must be identified that are accessible by a variational 
calculations.   

In particular, in this work we adopt a variational 
technique based on a recent multi-band extension~\cite{mio} of the so-called Gutzwiller approximation 
to evaluate analytically average values on Gutzwiller variational wavefunctions.~\cite{Gutzwiller1,Gutzwiller2} 
This method is not exact like VMC, unless in the case of infinite-coordination lattices. However, we have found that  
a variational wavefunction richer than that of Ref.~\onlinecite{Ogata} seems to compensate for the approximation 
adopted to calculate average values, thus leading to the same phase-diagram as the one obtained by VMC in the case of a two-dimensional 
square lattice~\cite{Ogata}. Encouraged by this result, we have extended the analysis of Ref.~\onlinecite{Ogata}. Specifically, 
we have derived the phase diagram forcing the wavefunction to remain paramagnetic. Similarly to the CDMFT calculation of 
Ref.~\onlinecite{DeLeo}, we have found that a transition accompanied by a topological change of the Fermi surface exists 
also in this case, although is masked by magnetism when we allow for it. Finally, we have analyzed the role of a uniform magnetic 
field in the paramagnetic phase and found a matamagnetic instability near the above phase transition, suggestive of the 
metamagnetism observed experimentally.~\cite{Flouquet}

The plan of the paper is as follows. In section II we introduce the model and the variational technique. The variational phase 
diagram is presented in section III. In section IV we discuss the properties of the variational wavefunction in the paramagnetic 
sector, while in section V we consider the effect of a magnetic field in connection with metamagnetism. Section VI is devoted 
to concluding remarks. Finally, in the appendix we present some technical details of the variational method employed.

\section{The Model and the variational method}

We consider a Kondo lattice model (KLM) described the Hamiltonian 
\bea
\mathcal{H} &=& -t\sum_{<\bR\bRp>}\,\sum_\sigma\, \bigg(c^\dagger_{\bR\sigma}c^\dagga_{\bRp\sigma}+H.c.\bigg) \nonumber \\
&& + J\,\sum_\bR\,\mathbf{S}_{f\bR}\cdot\mathbf{S}_{c\bR} \equiv \mathcal{H}_0 + \mathcal{H}_J,\label{Ham}
\eea
where $c^\dagger_{\bR\sigma}$ creates a conduction electron at site $\bR$ with spin $\sigma$ that can hop with amplitude 
$-t$ to nearest neighbor sites, $\mathbf{S}_{f\bR}$ is the spin-1/2 operator of the $f$-orbital and $\mathbf{S}_{c\bR}$ 
the conduction electron spin-density at site $\bR$. In what follows, we assume a bipartite lattice. To study this Hamiltonian 
we introduce the following variational Gutzwiller wavefunction
\be
|\Psi\rangle = \prod_\bR\,\mathcal{P}_\bR\,|\Psi_0\rangle,\label{GWF}
\ee
where $|\Psi_0\rangle$ is the ground state of a non-interacting two-band variational Hamiltonian describing hybridized 
$c$ and $f$ orbitals, while $\mathcal{P}_\bR$ is a local operator that modifies the relative weights of the local 
electronic configurations with respect to the uncorrelated wavefunction. In particular, we will assume 
for $\mathcal{P}_\bR$ the general expression
\be
\mathcal{P}_\bR = \sum_{\Gamma,n}\,\lambda_{\Gamma n}(\bR)\,\Pj{\Gamma,\bR}{n,\bR},\label{PR}
\ee
where $\ket{\Gamma,\bR}$ and $\ket{n,\bR}$ span all electronic configurations of the $c$ and $f$ orbitals at site $\bR$, with 
the constraint that the states $\ket{\Gamma,\bR}$, but not $\ket{n,\bR}$, have just a single $f$-electron.
 
The variational wavefunction \eqn{GWF} has been widely used to study the periodic Anderson model as well as 
its strong coupling counterpart, the Kondo lattice model, within the Gutzwiller 
approximation.~\cite{Brandow,Rice&Ueda,Varma,Abrahams,Fazekas1,Fazekas} However, in all the earlier works the operator   
$\mathcal{P}_\bR$ has been choosen to act only on the $f$-orbitals states. For instance, in the KLM that we consider, 
this choice would reduce to take $\mathcal{P}_\bR$ as the projector onto singly occupied $f$-orbitals, namely   
\be
\mathcal{P}_\bR = \sum_{\Gamma}\Pj{\Gamma,\bR}{\Gamma,\bR} = \left(n_{f\bR\up}-n_{f\bR\giu}\right)^2,\label{PR0}
\ee
where $n_{f\bR\sigma}= f^\dagger_{\bR\sigma}f^\dagga_{\bR\sigma}$. This assumption implies that the spin correlations 
induced by the exchange $J$ in \eqn{Ham} are only provided by the uncorrelated wavefunction $\ket{\Psi_0}$. The more general 
form of  $\mathcal{P}_\bR$, Eq.~\eqn{PR}, that we assume in what follows, permits to include additional 
correlations besides those included in the wavefunction $\ket{\Psi_0}$, in particular the tendency of the conduction electrons 
to couple into a singlet with the localized spins. 

The variational procedure amounts to optimize both the parameters $\lambda_{\Gamma n}(\bR)$ as well as those that 
identify $\ket{\Psi_0}$ by minimizing the average value of the Hamiltonian \eqn{Ham}. In general this task can be accomplished 
only numerically, for instance by means of VMC as actually done by Watanabe and Ogata~\cite{Ogata} 
with the simple choice of $\mathcal{P}_\bR$ as in Eq.~\eqn{PR0}. However, in infinite coordination lattices many simplications intervene 
that allow to evaluate average values analytically.~\cite{Metzner-Vollhardt-PRL,Metzner-Vollhardt-PRB,Gebhard} 
In this work we follow an extension~\cite{mio} of the multi-band method developed 
by B\"unemann, Weber and Gebhard~\cite{Gebhard,Bunemann} that allows to handle with non-hermitean operators 
$\mathcal{P}_\bR$, which is generally the case since the bra $\langle n,\bR|$ 
in \eqn{PR} can have any number of $f$-electrons while the ket $\ket{\Gamma,\bR}$ is forced to have only one. 

We start assuming that $\mathcal{P}_\bR$ is not the most general as possible but is subject to the following two 
conditions~\cite{nota1}
\bea
\Av{\Psi_0}{\mathcal{P}^\dagger_\bR\,\mathcal{P}^\dagga_\bR} &=& 1,\label{uno}\\
\Av{\Psi_0}{\mathcal{P}^\dagger_\bR\,\mathcal{P}^\dagga_\bR\,\mathcal{C}_{\bR\sigma}} &=& 
\Av{\Psi_0}{\mathcal{C}_{\bR\sigma}},\label{due}
\eea 
where 
\be
\mathcal{C}_{\bR\sigma} = 
\left(
\begin{array}{cc}
c^\dagger_{\bR\sigma}c^\dagga_{\bR\sigma} & c^\dagger_{\bR\sigma}f^\dagga_{\bR\sigma}\\
f^\dagger_{\bR\sigma}c^\dagga_{\bR\sigma} & f^\dagger_{\bR\sigma}f^\dagga_{\bR\sigma}\\
\end{array}
\right),\label{CR}
\ee
is the local single-particle density matrix operator. If Eqs.~(\ref{uno}) and (\ref{due}) are satisfied, then one can 
show~\cite{Gebhard,Bunemann,mio} that, in an infinite-coordination lattice, the average value of \eqn{Ham} 
that has to be minimized is 
\bea
E &=& \fract{\Av{\Psi}{\mathcal{H}}}{\langle \Psi|\Psi\rangle} \nonumber\\
&=& -t\sum_{<\bR\bRp>\sigma}\,\langle \Psi_0 |\,\Bigg[ 
\bigg(Z_{cc\sigma}(\bR)\,c^\dagger_{\bR\sigma} + Z_{cf\sigma}(\bR)\,f^\dagger_{\bR\sigma}\bigg)\nonumber \\
&&~~\bigg(Z^*_{cc\sigma}(\bRp)\,c^\dagga_{\bRp\sigma} + Z^*_{cf\sigma}(\bRp)\,f^\dagga_{\bRp\sigma}\bigg)
+ H.c.\Bigg]\,|\Psi_0\rangle
\nonumber\\
&& + J\sum_\bR\, \langle \Psi_0 |\,\mathcal{P}^\dagger_\bR\,\mathbf{S}_{f\bR}\cdot\mathbf{S}_{c\bR}\,
\mathcal{P}^\dagga_\bR\,|\Psi_0\rangle.\label{E-var}
\eea  
The hopping renormalization coefficients $Z$ are obtained through the following equations:
\begin{widetext} 
\bea
&&\langle \Psi_0 |\,\mathcal{P}^\dagger_\bR\, c^\dagger_{\bR\sigma}\,\mathcal{P}^\dagga_\bR\,
c^\dagga_{\bR\sigma}\,|\Psi_0\rangle =
Z_{cc\sigma}(\bR)\,\langle \Psi_0 | c^\dagger_{\bR\sigma}c^\dagga_{\bR\sigma}|\Psi_o\rangle 
+ Z_{cf\sigma}(\bR)\,\langle \Psi_0 | f^\dagger_{\bR\sigma}c^\dagga_{\bR\sigma}|\Psi_o\rangle \label{Z-cc}\\
&&\langle \Psi_0 |\,\mathcal{P}^\dagger_\bR\, c^\dagger_{\bR\sigma}\,\mathcal{P}^\dagga_\bR\,
f^\dagga_{\bR\sigma}\,|\Psi_0\rangle
= Z_{cc\sigma}(\bR)\,\langle \Psi_0 | c^\dagger_{\bR\sigma}f^\dagga_{\bR\sigma}|\Psi_o\rangle 
+ Z_{cf\sigma}(\bR)\,\langle \Psi_0 | f^\dagger_{\bR\sigma}f^\dagga_{\bR\sigma}|\Psi_o\rangle.\label{Z-cf} 
\eea
\end{widetext}
Therefore the variational calculation reduces, in infinite coordination lattices and provided 
Eqs.~(\ref{uno}) and (\ref{due}) are satisfied, to calculate expectation values on the 
Slater determinant uncorrelated wavefunction, which is analytically feasible since Wick's theorem applies.  
In the appendix we show how one can manipulate and simplify all the above expressions to get manageable formulas.
For finite-coordination lattices the expression \eqn{E-var} for the variational energy is nomore correct. However, it is 
common to keep using also in these cases the same formula \eqn{E-var} with the coefficients $Z$ defined above - the 
so-called Gutzwiller approximation~\cite{Gutzwiller1,Gutzwiller2}. Moreover, it is also common to intepret~\cite{Brandow,Gebhard-FL} 
the non-interacting Hamiltonian $\mathcal{H}_*$, see Eq.~\eqn{H-variational} in the appendix, 
whose ground state is the optimized wavefunction $|\Psi_0\rangle$, as the Hamiltonian of the quasiparticles within a 
Landau-Fermi liquid framework. 

Before moving to the presentation of our variational results, we want to mention some important consequences of 
choosing $\mathcal{P}^\dagga_\bR$ that acts both on the $f$ and on the $c$ orbitals. 
A drawback of the conventional Gutzwiller wavefunction with $\mathcal{P}^\dagga_\bR$ of Eq.~\eqn{PR0}, which was pointed 
out already by Fazekas and M\"uller-Hartmann in Ref.~\onlinecite{Fazekas}, is that, for small $J$, the paramagnetic solution 
gains a singlet-condensation energy that has a Kondo-like expression $\propto \exp(-1/J\rho)$, with 
$\rho$ the conduction electron density of states at the chemical potential. On the contrary, any magnetic solution 
gains a local exchange energy of order $J^2\rho$ - the average value of $J\sum_\bR\,\mathbf{S}_{f\bR}\cdot\mathbf{S}_{c\bR}$ - 
independently of the spatial arrangement of the magnetic ordering. This result would remain true even for a single impurity 
Kondo model and is obviously incorrect. Our wavefunction partially cures this deficiency because $\mathcal{P}_\bR$ is able to 
induce additional spin-correlations among $c$ and $f$ electrons, although only locally.   
   
We further note from \eqn{E-var} that the 
action of the Gutzwiller operator $\mathcal{P}^\dagga_\bR$ effectively generates an intersite hopping between the $f$-electrons, 
absent in the original Hamiltonian \eqn{Ham}, which correlates different sites hence can play an important role in 
determining the topology of the Fermi surface as well as in stabilizing magnetic structures.  Even though our method for 
computing average values is not exact in finite-coordination lattices, the more involved form of $\mathcal{P}^\dagga_\bR$ 
of Eq.~\eqn{PR} with respect to \eqn{PR0} partly compensates for this weakness -- the variational Hamiltonian 
contains inter-site $f$-$f$ and $f$-$c$ hopping -- leading to results that are very similar to 
those obtained by exact VMC, as we are going to show.

\section{Variational phase diagram}

We have solved the variational problem numerically using, for numerical convenience, a flat conduction-electron density-of-states 
with half-bandwidth $D$, our unit of energy. We do not expect that a more realistic density of states could qualitatively change  
the phase diagram that we find. Some technical details of the calculations are presented in the appendix, while here we just discuss 
the results. 
\begin{figure}
\begin{center}
\includegraphics[width=8cm]{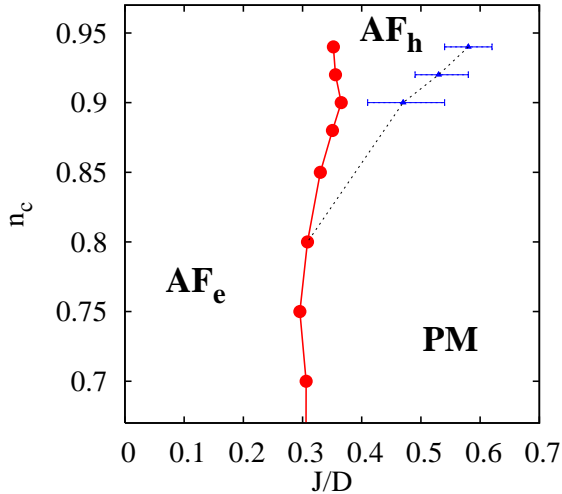}
\caption{\label{phd-AF}(Color online) Variational phase diagram as function of the conduction electron 
density $n_c$ and of the Kondo exchange in units of half the bandwidth, $J/D$. The solid line with the circles 
represents a first order line, while the dotted line is a second order transition. The error bars along the second order phase 
transition line reflect the variational uncertainty of a precise location of the continuous transition. 
The same problem does not arise along the discontinuous first order line. PM stands for paramagnetic 
heavy-fermion metal, while AF stands for an itinerant 
antiferromagnet, the subscripts ``e'' and ``h'' are borrowed from Ref.~\onlinecite{Ogata} and refer to the electron-like, ``e'', or 
hole-like, ``h'', character of the Fermi surface, see Fig.~\ref{FS-AF}.
}
\end{center}
\end{figure}

\begin{figure}
\begin{center}
\includegraphics[width=4cm]{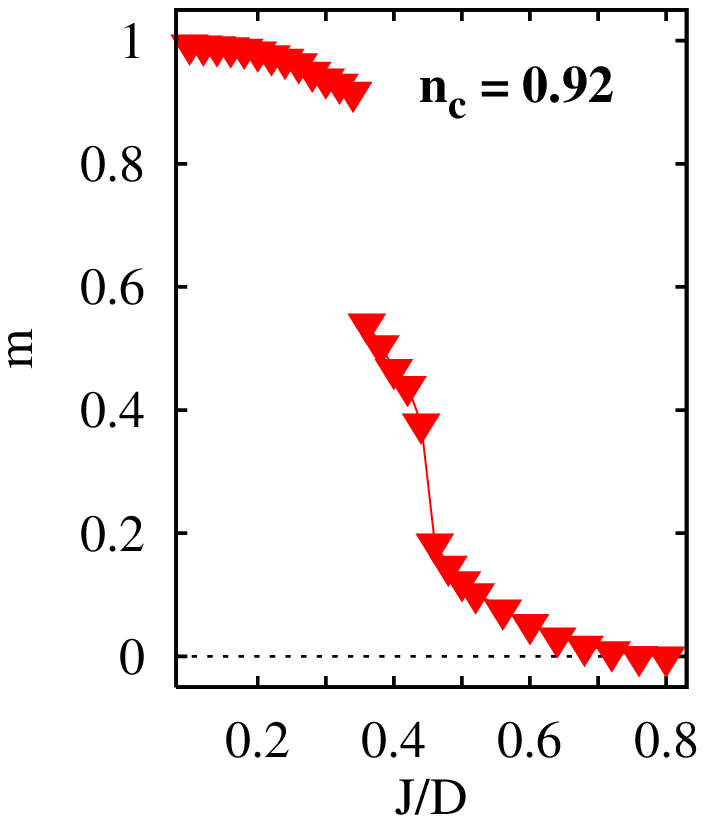}
\includegraphics[width=4cm]{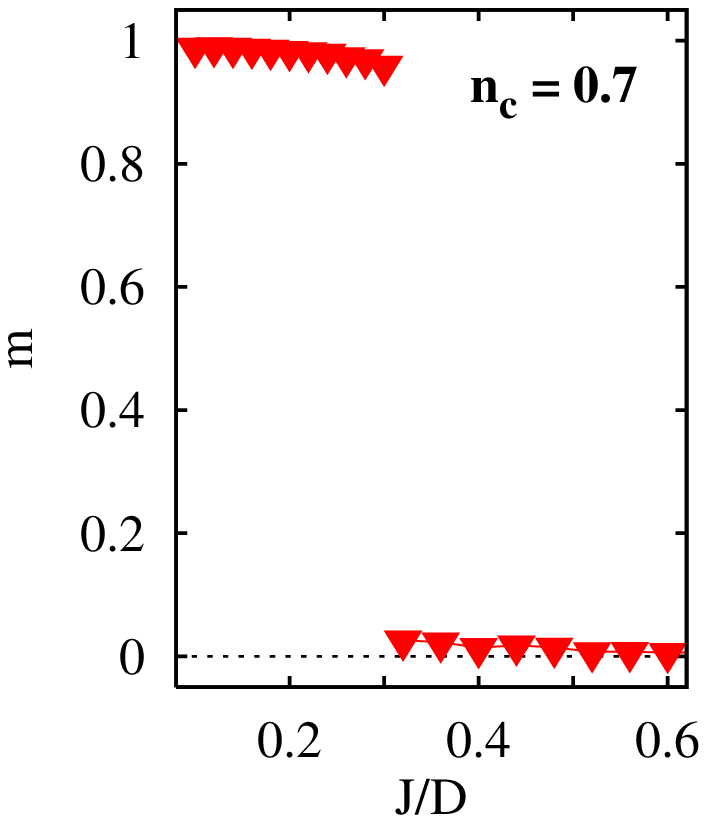}
\caption{\label{AF-order}(Color online) The magnetic order parameter as function of $J/D$ for $n_c=0.92$ (left panel)
and $n_c=0.7$ (right panel). Notice that for $n_c=0.92$ the order parameter grows continuously below a critical 
$J/D\simeq 0.6$ -- second order phase transition -- until at $J/D\simeq 0.36$ it jumps abruptly -- first order transition. 
For $n_c=0.7$ only a first order transition with a jump from zero to a finite value of the order parameter is observed.}
\end{center}
\end{figure}

\begin{figure}[thb]
\includegraphics[width=2.6cm]{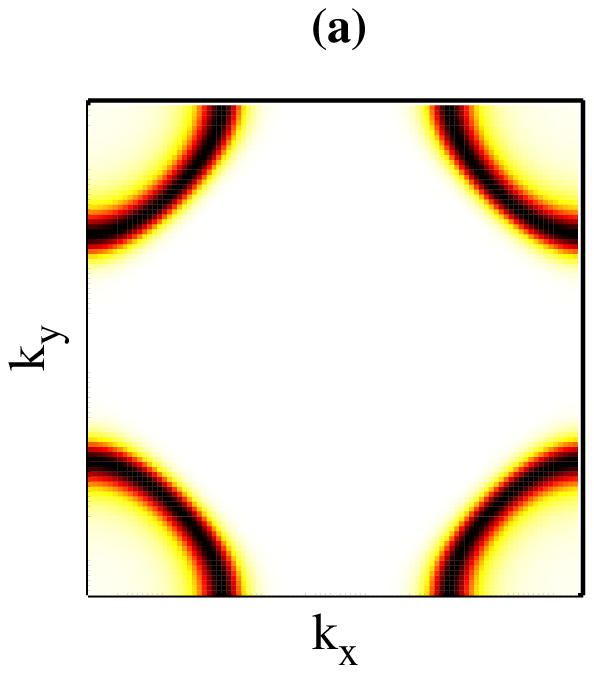}
\includegraphics[width=2.6cm]{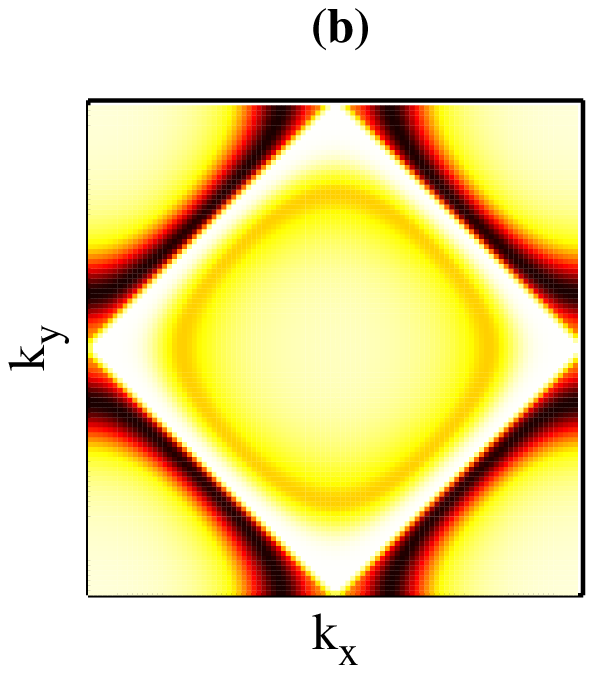}
\includegraphics[width=2.6cm]{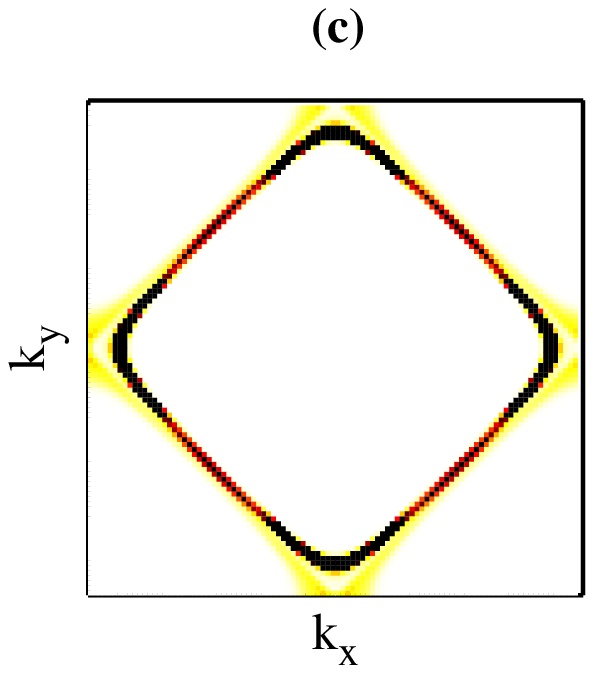}
\includegraphics[width=2.6cm]{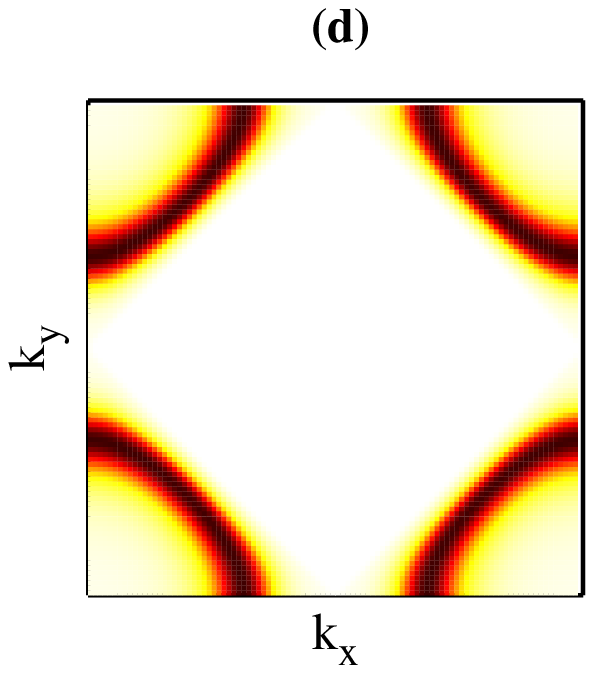}
\includegraphics[width=2.6cm]{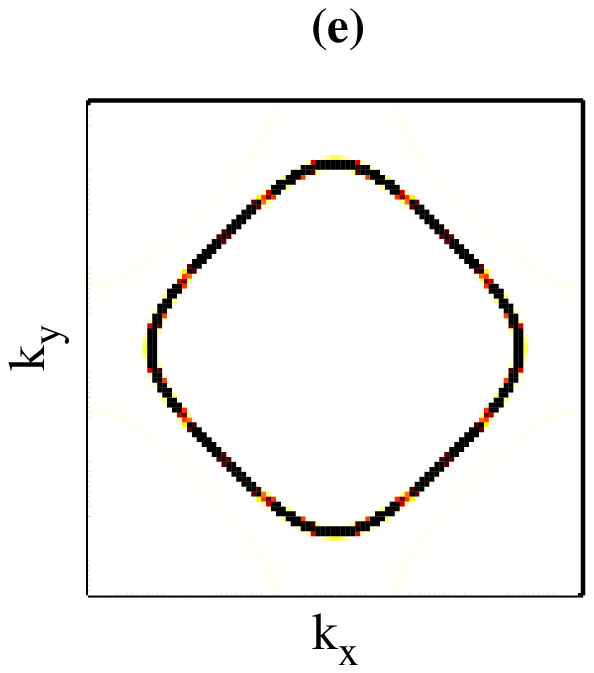}
\caption{\label{FS-AF}(Color online) The conduction electron spectral function at the chemical potential for a two-dimensional square lattice.  
Panels (a)-(b)-(c) show the evolution of the spectral function $A(\bk)$ at the chemical potential for $n_c= 0.92$ 
in the paramagnetic phase, panel (a) with $J/D=0.8$, 
right after the second-order transition, panel (b) with $J/D=0.4$, and finally below the first-order transition, panel (c) 
with $J/D=0.16$. Panels (d)-(e) show the same evolution with $n_c=0.7$ where there is only the first-order transition. 
}
\end{figure}

In Fig.~\ref{phd-AF} we show the variational phase diagram as function of the Kondo exchange $J$, in units of $D$, versus the 
conduction electron density $0\leq n_c < 1$. Close to the compensated regime $n_c=1$, one conduction electron per spin, 
we do find, similarly to Watanabe and Ogata~\cite{Ogata}, two successive transitions as $J/D$ is reduced from the heavy-fermion 
paramagnetic phase. First, N\`eel antiferromagnetism appears by a second order phase transition, see Fig.~\ref{AF-order}. 
Within the antiferromagnetic phase, a first-order phase transition further occurs at smaller $J/D$, see the 
jump of the order parameter in Fig.~\ref{AF-order}, accompanied by a 
rearrangement of the Fermi surface. This is shown in Fig.~\ref{FS-AF}, where we draw the quasiparticle (emission) spectral function 
at the chemical potential, defined by
\be
A(\bk) = -\int d\epsilon A(\bk,\epsilon)\,\fract{\partial f(\epsilon)}{\partial \epsilon},
\label{Ak}
\ee
where $f(\epsilon)$ is the Fermi distribution function at low temperature and 
\begin{widetext}
\bea
A(\bk,\epsilon) &=& \frac{\pi}{V}\sum_{\bR\bRp}\sum_\sigma\sum_n\,\mathrm{e}^{i\bk(\bR-\bRp)}
\,\delta\left(E_n-E_0-\epsilon\right)
\langle \Psi_0 |
\bigg(Z_{cc\sigma}(\bR)\,c^\dagger_{\bR\sigma} + Z_{cf\sigma}(\bR)\,f^\dagger_{\bR\sigma}\bigg)|\Psi_n\rangle\nonumber \\
&& \langle \Psi_n|\bigg(Z^*_{cc\sigma}(\bRp)\,c^\dagga_{\bRp\sigma} + Z^*_{cf\sigma}(\bRp)\,f^\dagga_{\bRp\sigma}\bigg)|\Psi_0\rangle
\label{Ake}
\eea
\end{widetext}
with $V$ the number of lattice sites. $|\Psi_0\rangle$ and $|\Psi_n\rangle$ are the ground state and an excited state, with 
energy $E_0$ and $E_n$, respectively, of the variational Hamiltonian $\mathcal{H}_*$, see Eq.~\eqn{H-variational} in the 
appendix. 
$A(\bk,\epsilon)$ is calculated with a nearest-neighbor hopping on a 
two dimensional square lattice, though with variational parameters optimized using a flat density of states at the same values of 
$n_c$ and $J/D$. 

The $\bk$-points where $A(\bk,\epsilon)$ is large identify the effective Fermi surface. 
We note that, in the paramagnetic phase, the Fermi surface is hole-like just as if the $f$ spins 
do partecipate the Luttinger sum rules - two bands with $1+n_c\leq 2$ electrons per site; one band empty and the other occupied 
by $1<1+n_c<2$ electrons.      
The same feature is also found beyond the second order phase transition. However, for $J/D$ below the first order phase transition, 
the Fermi surface changes topology 
and become electron-like, as if the $f$-electrons disappear from the Fermi surface.
Comparing the phase diagram Fig.~\ref{phd-AF} with the one obtained by VMC~\cite{Ogata}, we find that the two agree well, 
even quantitatively.~\cite{nota-Ogata} In order to identify the origin of the Fermi surface rearrangement, it is convenient to 
write the general expression of the variational Hamiltonian $\mathcal{H}_*$, see Eq.~\eqn{H-variational}, of which 
$|\Psi_0\rangle$ is the ground state. In momentum space and within the magnetic Brillouin zone
\begin{widetext}
\be
\mathcal{H}_* = \sum_\sigma\,\sum_{\bk\in MBZ}\, 
\psi^\dagger_{\bk\sigma}\,
\left(
\begin{array}{cccc}
t_{cc}\epsilon_\bk & V_u + t_{cf}\epsilon_\bk & \sigma m & \sigma V_s + \sigma t_{cf}'\epsilon_\bk \\
V_u + t_{cf}\epsilon_\bk  & \epsilon_f + t_{ff}\epsilon_\bk & \sigma V_s - \sigma t_{cf}'\epsilon_\bk & \sigma M \\
\sigma m & \sigma V_s - \sigma t_{cf}'\epsilon_\bk& - t_{cc}\epsilon_\bk & V_u - t_{cf}\epsilon_\bk  \\
\sigma V_s + \sigma t_{cf}'\epsilon_\bk & \sigma M & V_u - t_{cf}\epsilon_\bk  & \epsilon_f - t_{ff}\epsilon_\bk\\
\end{array}
\right)\,\psi^\dagga_{\bk\sigma},\label{H-var-explicit}
\ee
\end{widetext}
where $\epsilon_\bk$ is the energy dispersion of the conduction electrons, 
\[
\psi^\dagger_{\bk\sigma} = \left(c^\dagger_{\bk\sigma},f^\dagger_{\bk\sigma},c^\dagger_{\bk+\bQ\sigma},
f^\dagger_{\bk+\bQ\sigma}\right),
\]
a Fermi spinor, its hermitean conjugate being $\psi^\dagga_{\bk\sigma}$, $\bQ$ the N\`eel magnetic vector, 
and all the Hamiltonian parameters are variational but $\epsilon_\bk$. 

\begin{figure}
\begin{center}
\includegraphics[width=4cm]{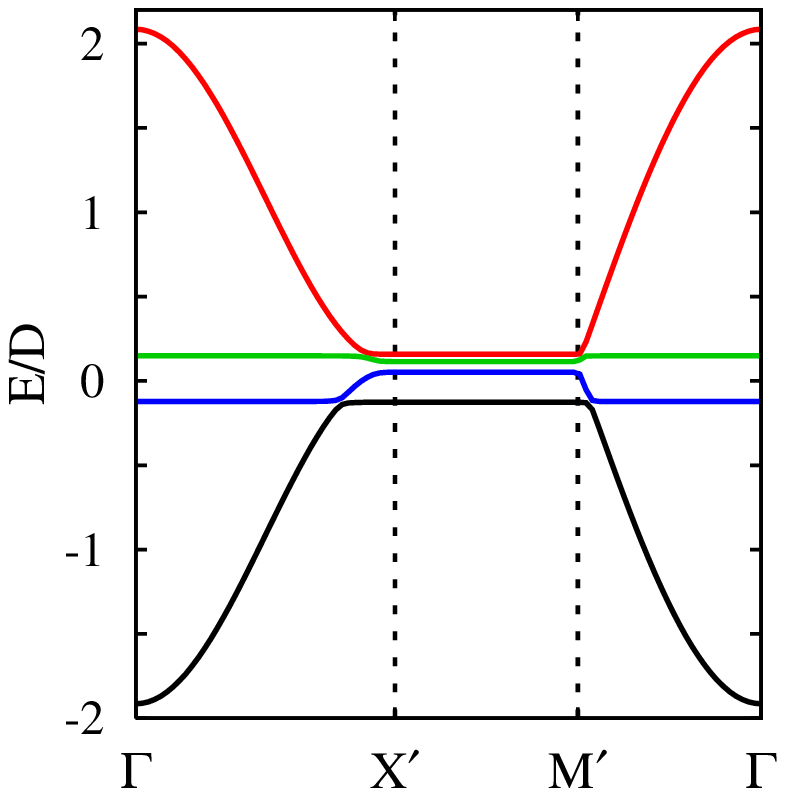}
\includegraphics[width=4cm]{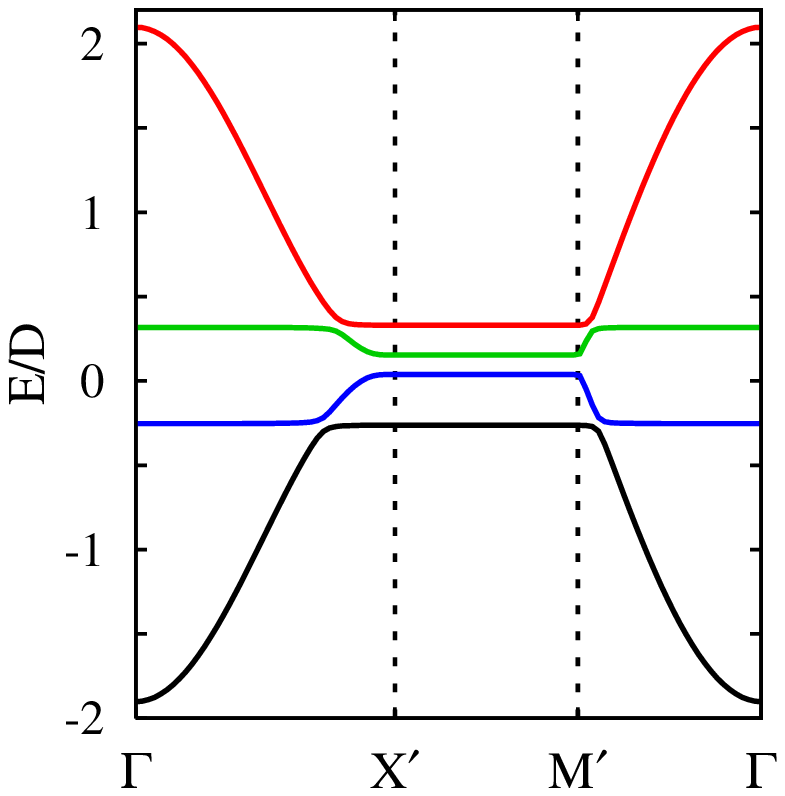}
\includegraphics[width=4cm]{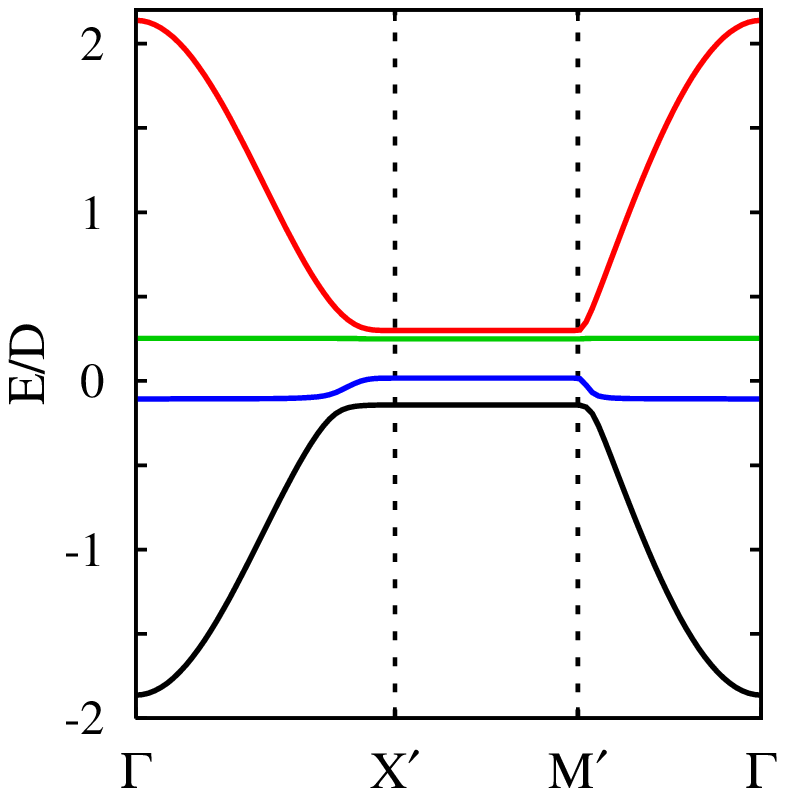}
\includegraphics[width=4cm]{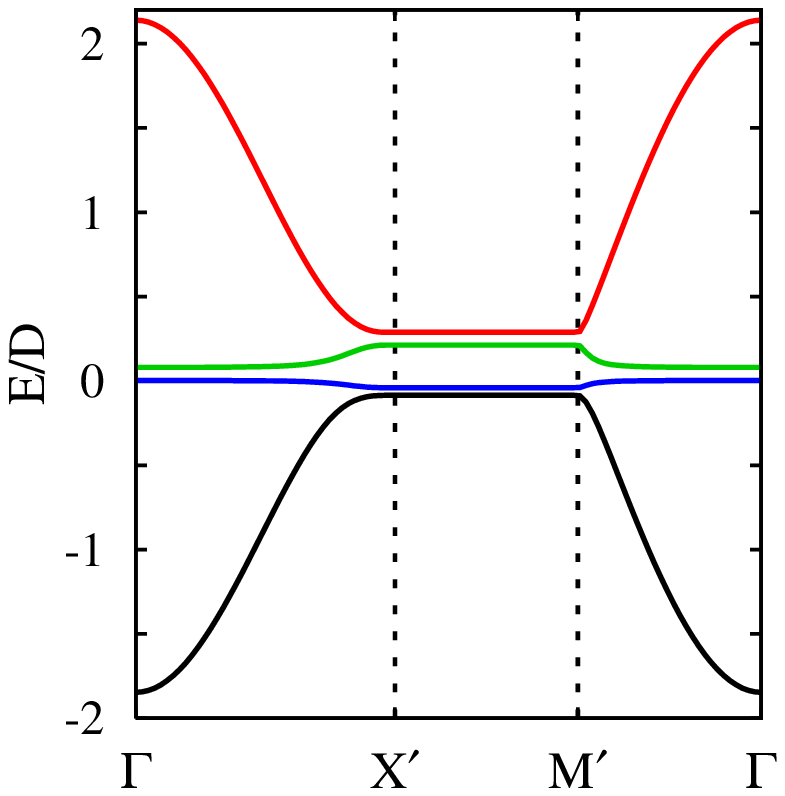}
\caption{\label{bande-AF} (Color online) Evolution of the band structure of the optimized variational Hamiltonian Eq.~\eqn{H-var-explicit} 
for $n_c=0.92$ as a function of $J/D$ and across the first order transition. From top left to bottom right panel: $J/D=0.1,\,J/D=0.2,\,J/D=0.3$ (below
the first-order transition) and $J/D=0.36$ (above the first-order transition). }
\end{center}
\end{figure} 

\begin{figure}
\includegraphics[width=8cm]{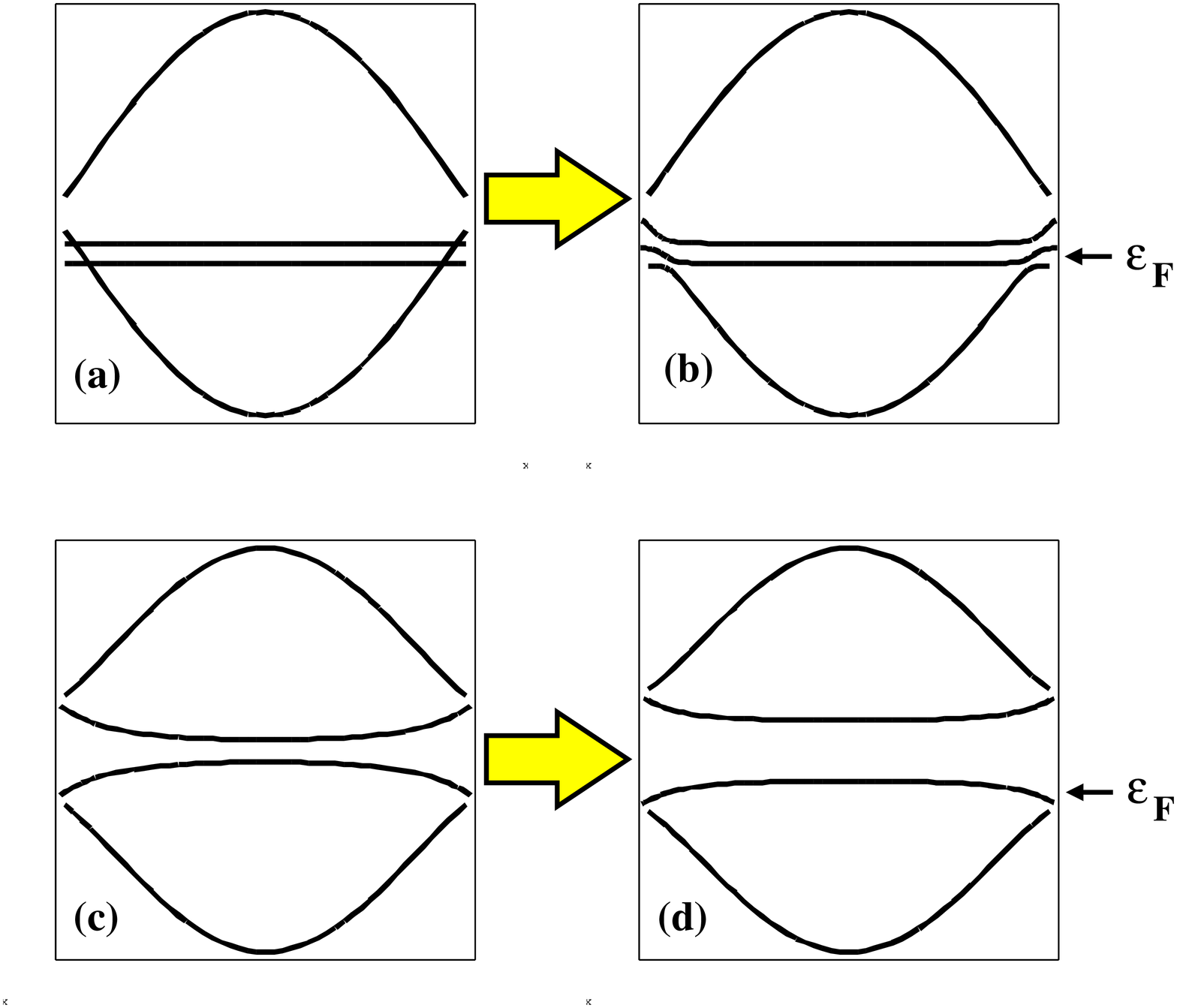}
\caption{\label{AF1}(Color online) One-dimensional representation of the different variational band structures 
in the two magnetic phases close to $n_c=1$, drawn in the magnetic Brillouin zone. Small $J/D$ phase: panel (a) 
represents non-hybridized $c$ and $f$ bands split by antiferromagnetism; panel (b) what happens once a small 
hybridization is switched on. Large $J/D$ phase: panel (c) represents non-magnetic hybridized $c$ and $f$ bands in the 
folded Brillouin zone; panel (d) what happens once a small antiferromagnetic order parameter is switched on. 
}
\end{figure}

In Fig.~\ref{bande-AF} we plot the variational bands in the antiferromagnetic 
phase below and above the first order phase transition. In agreement with the interpretation given by Watanabe and Ogata in  
Ref.~\onlinecite{Ogata}, the bands in the antiferromagnetic phase at low $J/D$   
can be sought as antiferromagnetically split $c$ and $f$ bands very weakly hybridized, panels (a) and (b) in Fig.~\ref{AF1}, 
while those at larger $J/D$ as strongly hybridized $c$ and $f$ bands weakly antiferromagnetically split, 
panels (c) and (d) in Fig.~\ref{AF1}. The main control parameter of the transition is the relative strength of 
the $f$-orbital energy, $\epsilon_f$ in \eqn{H-var-explicit}, with respect to the antiferromagnetic splittings, mostly $\sigma M$ 
in \eqn{H-var-explicit}.

Above a critical doping away from the compensated regime, we only find a single first-order phase transition transition, 
see Fig.~\ref{AF-order}, directly from a paramagnet at large $J/D$, with a band structure similar 
to panel (c) in Fig.~\ref{AF1} unfolded in the whole 
Brillouin zone, to an antiferromagnet with a band structure similiar to panel (b) in Fig.~\ref{AF1}. In other 
words, this phase transition is accompanied by a drastic reconstruction of the Fermi surface.

\section{Fermi-surface reconstruction vs. magnetism}

The variational phase diagram, Fig.~\ref{phd-AF}, shows that the onset of magnetism is not necessarily accompanied by 
a Fermi surface reconstruction. Viceversa, one could speculate that the latter might not require magnetism, which would be the case 
if the Fermi-surface change were caused by the $f$-electron localization~\cite{Johansson}.
This aspect makes worth investigating the properties of the variational wavefunction \eqn{GWF} preventing antiferromagnetism, 
which amounts to assume $\lambda_{\Gamma n}(\bR)$ in Eq.~\eqn{PR} independent of $\bR$ and $|\Psi_0\rangle$ a paramagnetic 
Slater determinant. 

\begin{figure}
\begin{center}
\includegraphics[width=8cm]{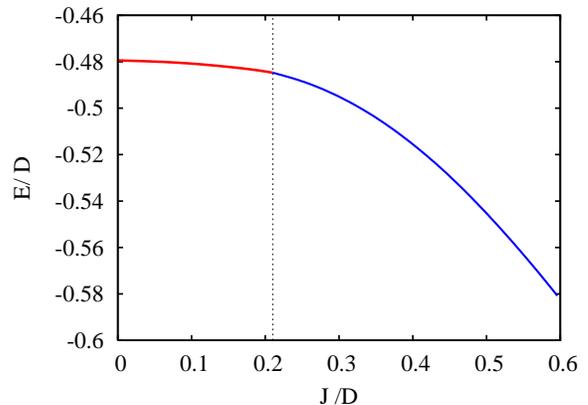}
\caption{\label{E-PARA}(Color online) Variational energy as function of $J/D$ at $n_c=0.8$ in the paramagnetic sector. 
A kink is visible at $J/D\simeq 2.1$. We note the finite curvature of the energy at low $J/D$, which, as we checked, is compatible 
with second order perturbation theory.}
\end{center}
\end{figure}

\begin{figure}
\includegraphics[width=4.2cm]{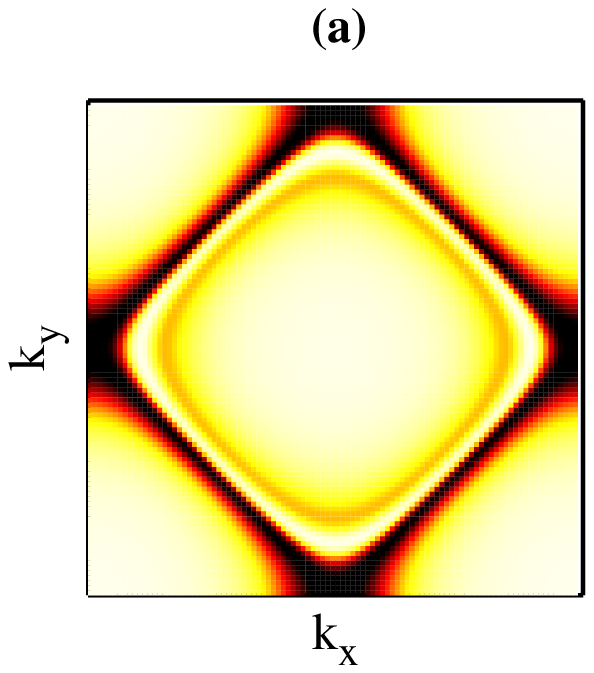}
\includegraphics[width=4.2cm]{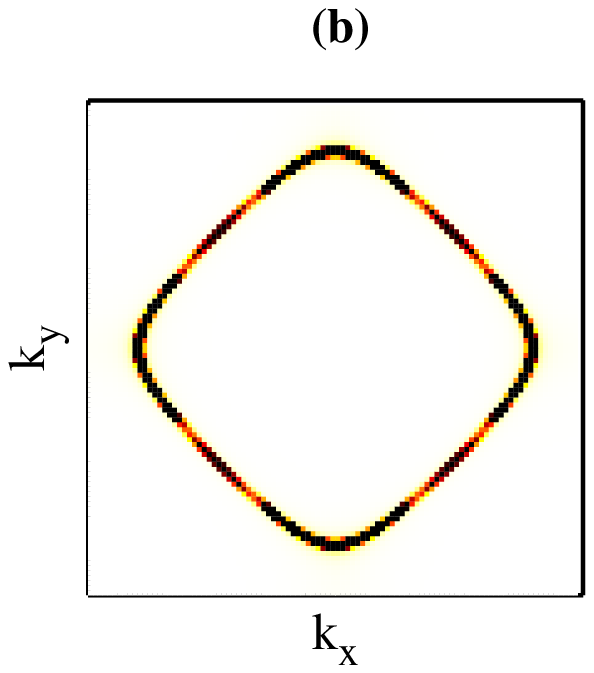}
\caption{\label{FS-PARA}(Color online) Spectral function at the chemical potential for $n_c=0.8$ and 
$J/D$ above, panel (a), and below, panel (b), the critical value.}
\end{figure} 

At first sight, one would not expect to find anything special varying $J/D$ in the paramagnetic sector. In fact, we previously 
mentioned that the change of the Fermi surface within the magnetic phase reflects essentially the change 
of the band structure, which, in turn, depends variationally only on the value of the $f$-orbital energy with respect to 
the magnetic splitting, respectively $\epsilon_f$ and $2M$ in Eq.~\eqn{H-var-explicit}.     
Therefore, without magnetism, i.e. $M=0$, the topology of the band structure must remain invariant whatever $J/D\not = 0$ is, 
as we indeed find. Nevertheless, even in this case, we do observe a very weak first order phase transition for values of $J/D$ 
slightly smaller than those at which the first order transition occurs when we allow for magnetism, as shown by the behavior of the 
variational energy in Fig.~\ref{E-PARA}. Remarkably, at this transition the momentum dependent spectral 
function of the conduction electrons at the chemical potential  
changes abruptly, see Fig.~\ref{FS-PARA}. For $J/D$ above the critical value, the Fermi surface 
includes the $f$-electrons, while, below, it does not, exactly as we find when magnetism is present. 
However, this change occurs now not because the band structure is modified 
but because the spectral weight of the conduction electrons at the Fermi energy changes discontinously. Indeed, looking carefully 
at the momentum distribution in Fig.~\ref{FS-PARA}a, one can distinguish two sheets 
of the Fermi surface, a small one, which corresponds to 
the non-interacting conduction electron Fermi surface, and a large one that includes also the $f$ electrons. 
Across the transition, it is the relative weight of these two sheets that change discontinuously. 
We believe that this must be regarded as a manifestation of an $f$-localization, or, better, of an orbital-selective localization, 
as proposed in Refs.~\onlinecite{DeLeo} and \onlinecite{deleo-2008}.
This result also demonstrates that the rearrangement of the Fermi surface observed along the first-order 
line in the phase diagram Fig.~\ref{phd-AF} is caused by the $f$-electron orbital-selective localization rather than by magnetism.

\begin{figure}
\begin{center}
\includegraphics[width=8cm]{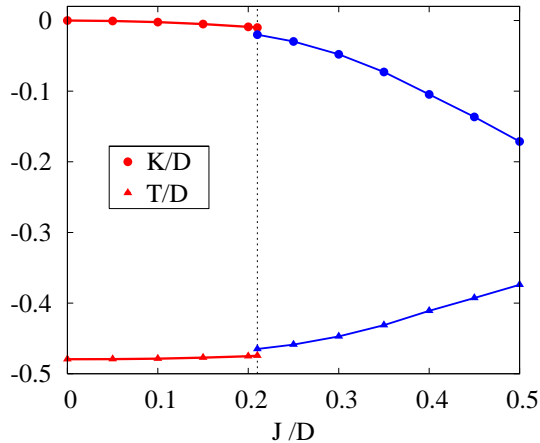}
\caption{\label{E-Kondo-hopping}(Color online) Behavior of the average Kondo exchange, $K/D$, and hopping, $T/D$ (in units of $D$, half the 
conduction bandwidth). }
\end{center}
\end{figure}

Inspection of the behavior of the average Kondo exchange and hopping, Fig.~\ref{E-Kondo-hopping}, shows that the 
``localized'' phase has a better conduction-electron hopping energy, while the ``delocalized'' one a better Kondo exchange. This 
suggests that the abrupt change of the Fermi surface is primarily consequence of the competition between the 
conduction electron band-energy and the Kondo exchange, and not of the commonly invoked competion between Kondo and RKKY interactions.
 
In light of these results, also the transition lines in the phase diagram, Fig.~\ref{phd-AF}, assume a different meaning. 
The first-order line that separates the paramagnet from the antiferromagnet is primarily due to the $f$-localization, magnetism being 
just its by-product. On the contrary, the second-order line close to the compensated regime 
is more likely to be interpreted as a Stoner's instability of the paramagnetic 
Fermi-liquid, driven by the nesting property of the Fermi surface at $n_c=1$. Across this second-order phase transition, the 
Fermi surface changes, smoothly, following the spin splitting of the bands.

\section{Metamagnetism}

Another indirect signal of the $f$-localization can be found by studying the behavior of the paramagnet in the presence of a uniform 
magnetic field. Indeed, if the $f$-orbitals are close to a Mott localization, they are also very prompt to order magnetically. 
Let alone, they would prefer some magnetic order along with the structure of the RKKY exchange, 
in our bipartite lattice model not far from half-filling the 
natural candidate being a N\'eel ordering. However, in the presence of a magnetic field, they could equally prefer to 
order ferromagnetically. 
In other words, it is plausible to foresee that the $f$-localization could be driven by a weak magnetic field, the weaker the closer 
the orbital-selective Mott transition is, thus accompanied by a sharp increase of magnetization, so-called metamagnetism, as well as by a 
discontinuous change of the Fermi surface. 

\begin{figure}
\begin{center}
\includegraphics[width=8cm]{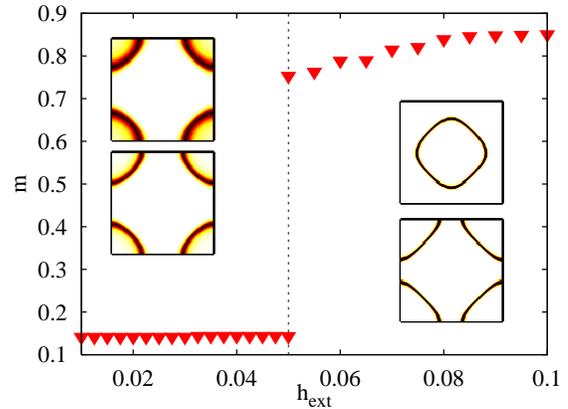}
\caption{\label{metamagnetism}(Color online) Evolution of the uniform magnetization as function of an external magnetic field applied in the
paramagnetic phase ($J/D=0.45$ at $n_c=0.88$). Insets show the spectral functions for the majority (top panel) and 
minority (bottom panel) spins across the metamagnetic transition.}
\end{center}
\end{figure}

This expectation is confirmed by our variational calculation. In 
Fig.~\ref{metamagnetism} we show the evolution of the uniform magnetization as function of the applied magnetic field in the 
paramagnetic phase at $J/D=0.45$ and $n_c=0.88$. Indeed, as function of the magnetic field, we do find a first order phase transition 
that is accompanied by a abrupt increase of the magnetization as well as by a discontinuous change of the conduction electron 
Fermi surface, specifically of the majority spin one. In fact, since the critical field is smaller than the Kondo exchange $J$, once the 
$f$ electrons localize and their spins align with the external field, the effective Zeman field felt by the conduction electrons 
is opposite to the applied one. Consequently, the Fermi surface of the majority spin becomes smaller than the minority spin one, 
contrary to the case for external fields below the metamagnetic transition, which is what we find, although hardly visible in 
Fig.~\ref{metamagnetism}. 

\section{Conclusions}

We have calculated within the Gutzwiller approximation the phase diagram of the Kondo lattice model as function of the conduction 
electron density and of the Kondo exchange $J$. The novel feature of our approach 
with respect to earlier ones is that the Gutzwiller projector acts on all the electronic configurations of each $f$ orbital 
plus the conduction state to which it is hybridized. This allows to include additional local correlations between $f$ and conduction 
electrons, specifically those that favour singlet pairing among them. Summarizing our variational results, we have found that:  
\begin{itemize}
\item there exists an orbital selective Mott localization of the $f$ electrons accompanied by a discontinuous change of the 
Fermi surface;
\item away from any nesting instability, this first-order transition in accompanied by magnetism; 
\item on the contrary, when the conduction electron Fermi surface is perfectly or almost perfectly nested, magnetism occurs before the 
$f$-localization, via a second order transition with a continuous change of the Fermi surface;
\item the $f$-electron Mott localization can be also induced by a uniform magnetic field, in which case it is revealed by a 
metamagnetic transition at which the magnetization jumps and the Fermi surface changes discontinuously. 
\end{itemize}
These findings bridge between the cluster dynamical mean field theory results of Refs.~\onlinecite{DeLeo}-\onlinecite{deleo-2008} 
and the variational Monte Carlo ones of Ref.~\onlinecite{Ogata}, and suggest that generically, i.e. without nesting, 
magnetism is a by-product of the $f$-electron Mott localization rather than the outcome of the competition between Kondo screening 
and RKKY interaction. We must mention that the weak first-order character of the Mott transition that we find might be a spurious 
outcome of the variational procedure, so that we can not exclude that in reality such a transition is continuous.    

The question we can not address, since ours is a variational approach for the ground state, concerns the anomalous thermodynamic 
behavior observed around the magnetic transition. In other words, we can not establish whether such a behavior is associated with 
the incipient magnetism~\cite{millis-1993} or is just a consequence of the $f$-electron 
localization,~\cite{Coleman-Pepin,Si} or better of the orbital selective Mott 
localization.~\cite{Pepin,Pepin-2,biermann,deleo-2008}

\begin{acknowledgments}
We thank Michel Ferrero for helpful discussions. 
\end{acknowledgments}

\appendix*
\section{The Gutzwiller approximation in detail}
%\label{Appendice1}
In this appendix we present some technical details of the method that we have employed, which simplify considerably all calculations.  
We start by the definition of the Gutzwiller wavefunction Eq.~\eqn{GWF} with the general Gutzwiller operator 
$\mathcal{P}_\bR$ of Eq.~\eqn{PR}. We assume that the 
average value of the local single-particle density-matrix operator, $\mathcal{C}_{\bR\sigma}$ in Eq.~\eqn{CR}, 
on the uncorrelated Slater determinant wavefunction $\Psi_0$ 
is diagonal in terms of the operators $d^\dagger_{1\bR\sigma}$ and $d^\dagger_{2\bR\sigma}$, related by a unitary transformation 
to the original ones, $c^\dagger_{\bR\sigma}$ and $f^\dagger_{\bR\sigma}$. In other words, for $a,b=1,2$, 
\be
\Av{\Psi_0}{d^\dagger_{a\bR\alpha}d^\dagga_{b\bR\beta}} = \delta_{ab}\,\delta_{\alpha\beta}\,n^0_{a\bR\alpha},
\label{C-av}
\ee
where $0\leq n^0_{a\bR\alpha}\leq 1$ are the eigenvalues of $\mathcal{C}_{\bR\sigma}$. 
We specify $\mathcal{P}_\bR$ to be of the form as in Eq.~\eqn{PR}, namely 
\be
\mathcal{P}_\bR = \sum_{\Gamma n}\,\lambda_{\Gamma n}(\bR)\, \Pj{\Gamma,\bR}{n,\bR},\label{PR-bis}
\ee
where the states $|\Gamma,\bR\rangle$ are written in the original $c$-$f$ basis and contain only one $f$ electron   
while, by assumption, $|n,\bR\rangle$ are Fock states in the natural basis, namely in terms of $d_1$-$d_2$. In other words, 
and dropping for simplicity the site-label $\bR$, a generic state $|n\rangle$ is identified by the occupation numbers 
$n_{a\sigma}=0,1$, $a=1,2$ and $\sigma=\su,\giu$, and has the explicit expression 
\[
|n\rangle = \left(d^\dagger_{1\up}\right)^{n_{1\up}}
\left(d^\dagger_{1\giu}\right)^{n_{1\giu}}\left(d^\dagger_{2\up}\right)^{n_{2\up}}
\left(d^\dagger_{2\giu}\right)^{n_{2\giu}}\,|0\rangle.
\]
We introduce the uncorrelated occupation-probability matrix $P^0$ with elements  
\be
P^0_{nm} \equiv \Av{\Psi_0}{|m\rangle\langle n|} = \delta_{nm}\,P^0_n,\label{P0}
\ee
where 
\be
P^0_n = \prod_{a=1,2}\,\prod_{\sigma=\su,\giu}\, \left(n^0_{a\sigma}\right)^{n_{a\sigma}}\,
\left(1-n^0_{a\sigma}\right)^{1-n_{a\sigma}}.\label{Pn0}
\ee
We also introduce the matrix representation of the operators $d^\dagga_{a\sigma}$ and $d^\dagger_{a\sigma}$, namely 
\ba
&&d^\dagga_{a\sigma}\rightarrow \left(d^\dagga_{a\sigma}\right)_{nm} = \langle n|d^\dagga_{a\sigma}|m\rangle,\\
&&d^\dagger_{a\sigma}\rightarrow \left(d^\dagger_{a\sigma}\right)_{nm} = \langle n|d^\dagger_{a\sigma}|m\rangle
= \left(\langle m|d^\dagga_{a\sigma}|n\rangle\right)^*,
\ea
and assume that the variational parameters $\lambda_{\Gamma n}$ in Eq.~\eqn{PR-bis} are the elements of a matrix $\lambda$. 
With the above definitions, the two conditions Eqs.~\eqn{uno} and \eqn{due} that we impose, and which allow for an analytical 
treatment in infinite-coordination lattices, become~\cite{mio}   
\bea
&&\Av{\Psi_0}{\mathcal{P}^\dagger\,\mathcal{P}^\dagga} = 
\Tr\left(P^0\,\lambda^\dagger\,\lambda^\dagga\right) \nonumber\\
&& = \sum_{\Gamma n}\,P^0_n\,\lambda^\dagger_{n\Gamma}\,\lambda^\dagga_{\Gamma n} = 
1,\label{uno-app}\\
&&\Av{\Psi_0}{\mathcal{P}^\dagger\,\mathcal{P}^\dagga\,d^\dagger_{a\alpha}d^\dagga_{b\beta}} = 
\Tr\left(P^0\,\lambda^\dagger\,\lambda^\dagga\,d^\dagger_{a\alpha}
d^\dagga_{b\beta}\right)\nonumber\\
&& = \sum_{\Gamma n m}\,P^0_n\,\lambda^\dagger_{n\Gamma}\lambda^\dagga_{\Gamma m}\,
\langle m|d^\dagger_{a\alpha}d^\dagga_{b\beta}|n\rangle \nonumber\\
&& = \Av{\Psi_0}{d^\dagger_{a\alpha}d^\dagga_{b\beta}} = \delta_{ab}\,\delta_{\alpha\beta}\,n^0_{a\alpha}.\label{due-app}
\eea      
If Eqs.~\eqn{uno-app} and \eqn{due-app} are satisfied, then the average value of any local operator $\mathcal{O}$ in 
infinite-coordination lattices is~\cite{Gebhard,mio}
\bea
&&\Av{\Psi}{\mathcal{O}} = \Av{\Psi_0}{\mathcal{P}^\dagger\,\mathcal{O}\,\mathcal{P}}\nonumber\\
&&= \Tr\left(P^0\,\lambda^\dagger\,O\,\lambda\right) = \sum_{n\Gamma\Gamma'} P^0_n\, \lambda^\dagger_{n\Gamma}\, O_{\Gamma\Gamma'}\,
\lambda^\dagga_{\Gamma' n},\label{av-O}
\eea
where $O$ is a matrix with elements
\[
O_{\Gamma\Gamma'} = \langle \Gamma|\mathcal{O}|\Gamma'\rangle.
\]  
In the mixed original-natural basis representation, the proper definition of the $Z$-factors in Eqs.~\eqn{Z-cc} and \eqn{Z-cf} 
changes into  
\bea
&&\Av{\Psi_0}{\mathcal{P}^\dagger\,c^\dagger_\sigma\,\mathcal{P}\,d^\dagga_{1\sigma}} = 
Z_{c1\sigma}\,\Av{\Psi_0}{d^\dagger_{1\sigma}d^\dagga_{1\sigma}} \nonumber\\
&&~~~~~~~~~~+ Z_{c2\sigma}\,\Av{\Psi_0}{d^\dagger_{2\sigma}d^\dagga_{1\sigma}} \nonumber\\
&&~~= Z_{c1\sigma}\,n^0_{1\sigma}\label{Z-c1}\\
&&\Av{\Psi_0}{\mathcal{P}^\dagger\,c^\dagger_\sigma\,\mathcal{P}\,d^\dagga_{2\sigma}} = 
Z_{c1\sigma}\,\Av{\Psi_0}{d^\dagger_{1\sigma}d^\dagga_{2\sigma}} \nonumber\\
&&~~~~~~~~~~+ Z_{c2\sigma}\,\Av{\Psi_0}{d^\dagger_{2\sigma}d^\dagga_{2\sigma}} \nonumber\\
&&~~= Z_{c2\sigma}\,n^0_{2\sigma}.\label{Z-c2}
\eea
In other words, when calculating the average hopping, the operator $c^\dagger_\sigma$ effectively 
transforms into  
\be
c^\dagger_\sigma \rightarrow Z_{c1\sigma}\,d^\dagger_{1\sigma} + Z_{c2\sigma}\,d^\dagger_{2\sigma}.\label{Z-qp}
\ee

\subsection{Explicit formulas and connection with slave-boson mean field theory} 

To further simplify the calculation, we introduce a new matrix in the mixed basis representation 
\be
\phi = \lambda\,\sqrt{P^0},\label{phi}
\ee
with elements
\be
\phi_{\Gamma,n} = \lambda_{\Gamma n}\,\sqrt{P^0_n}.\label{phi_Gamma-n}
\ee
As we shall see, $\phi_{\Gamma n}$ corresponds to the slave-boson saddle-point value within the multiband extension of 
the Kotliar-Ruckenstein mean-field scheme recently introduced by Lechermann and coworkers~\cite{Georges}, which they named 
rotationally invariant slave-boson formalism. By means of the definition \eqn{phi} the first condition \eqn{uno} 
becomes 
\[
\Tr\left(\phi^\dagger \,\phi\right) = \sum_{\Gamma n}\,\phi^\dagger_{n\Gamma}\,\phi^\dagga_{\Gamma n}= 1,
\] 
which coincides with the saddle-point value of Eq.~(28) in Ref.~\onlinecite{Georges}. The other condition, Eq.~\eqn{due-app}, 
becomes 
\bea
&&\Tr\left(\sqrt{P^0}\,\phi^\dagger\,\phi\,\sqrt{\fract{1}{P^0}}\;d^\dagger_{a\alpha}
d^\dagga_{b\beta}\right) = \Av{\Psi_0}{d^\dagger_{a\alpha}d^\dagga_{b\beta}}\nonumber\\
&& = \sum_{\Gamma n m} \sqrt{\fract{P^0_n}{P^0_m}}\; \phi^\dagger_{n\Gamma}\phi^\dagga_{\Gamma m}\, 
\langle m|d^\dagger_{a\alpha}d^\dagga_{b\beta}|n\rangle\nonumber\\
&&= \delta_{ab}\,\delta_{\alpha\beta}\,n^0_{a\alpha},\label{intermediate}
\eea
which is apparently different from the saddle-point value of Eq.~(29) in Ref.~\onlinecite{Georges}, which reads 
\bea 
&&\Tr\left(\phi^\dagger\,\phi\,d^\dagger_{a\alpha}
d^\dagga_{b\beta}\right) = \sum_{\Gamma n m} \phi^\dagger_{n\Gamma}\phi^\dagga_{\Gamma m}\, 
\langle m|d^\dagger_{a\alpha}d^\dagga_{b\beta}|n\rangle\nonumber\\ 
&& = \Av{\Psi_0}{d^\dagger_{a\alpha}d^\dagga_{b\beta}}.\label{intermediate-georges}
\eea
Therefore the equivalence between the rotationally invariant slave-boson formalism and the multi-band Gutzwiller approximation 
is not so immediate as claimed recently by B\"unemann and Gebhard.~\cite{quaquaraqua} Indeed the proof given by these authors 
suffers by a flaw~\cite{sasso}, eventhough, as we are going to show, the final conclusion seems to be right. 
In fact, we note that the two Fock states $|n\rangle$ and $|m\rangle$ in Eq.~\eqn{intermediate} differ only because 
$|n\rangle$ has the orbital $b$ with spin $\beta$ occupied but orbital $a$ with spin $\alpha$ empty, while it is viceversa for $|m\rangle$, 
so that 
\[
\sqrt{\fract{P^0_n}{P^0_m}} = \sqrt{\fract{n^0_{b\beta}\left(1-n^0_{a\alpha}\right)}
{\left(1-n^0_{b\beta}\right)n^0_{a\alpha}}},
\]
hence Eq.~\eqn{intermediate} is actually equal to 
\begin{widetext}
\[
\sqrt{\fract{n^0_{b\beta}\left(1-n^0_{a\alpha}\right)}
{\left(1-n^0_{b\beta}\right)n^0_{a\alpha}}}\,\sum_{\Gamma n m} \, 
\phi^\dagger_{n\Gamma}\phi^\dagga_{\Gamma m}\, \langle m|d^\dagger_{a\alpha}d^\dagga_{b\beta}|n\rangle 
= \delta_{ab}\,\delta_{\alpha\beta}\,n^0_{a\alpha}.
\]
\end{widetext}
Because of $\delta_{ab}\delta_{\alpha\beta}$ on the r.h.s this equation is equivalent to Eq.~(\ref{intermediate-georges}) provided 
$n^0_{b\beta}\not = 0$ and $n^0_{a\alpha}\not = 1$. Therefore, if the average value of the single-particle density matrix on 
the uncorrelated wavefunction $|\Psi_0\rangle$ has eigenvalues neither 0 nor 1, the conditions Eqs.~\eqn{uno-app} and 
\eqn{due-app} are equivalent to impose 
\bea
\Tr\left(\phi^\dagger\,\phi\right) &=& 1,\label{uno-app-bis}\\
\Tr\left(\phi^\dagger\,\phi \,d^\dagger_{a\alpha}d^\dagga_{b\beta}\right) &=& 
\Av{\Psi_0}{d^\dagger_{a\alpha}d^\dagga_{b\beta}}\nonumber\\
&=& \delta_{ab}\,\delta_{\alpha\beta}\,n^0_{a\alpha},\label{due-app-bis}
\eea
which indeed coincide with Eqs.~(28) and (29) in Ref.~\onlinecite{Georges} specialized to the natural orbital basis and evaluated at 
the saddle point. 

In terms of $\phi$, the average of the local operator $\mathcal{O}$, Eq.~\eqn{av-O}, becomes
\be
\Av{\Psi}{\mathcal{O}} = \Tr\left(\phi^\dagger\, O\,\phi\right),\label{av-O-final}
\ee
which coincide with Eq.~(47) in Ref.~\onlinecite{Georges} at the saddle point. Finally we need to evaluate 
$Z_{c1\sigma}$ and $Z_{c2\sigma}$ of Eqs.~\eqn{Z-c1} and \eqn{Z-c2}. We find that, for $a=1,2$,  
\begin{widetext}
\bea
Z_{ca\sigma} &=& \fract{1}{n^0_{a\sigma}}\,\Tr\left(\sqrt{P^0}\,\phi^\dagger\,c^\dagger_\sigma\,\phi\,
\sqrt{\fract{1}{P^0}}\;d^\dagga_{a\sigma}\right) = 
\fract{1}{n^0_{a\sigma}}\,\sum_{\Gamma\Gamma' n m}\,\sqrt{\fract{P^0_n}{P^0_m}}\, 
\phi^\dagger_{n\Gamma}\,\langle \Gamma|c^\dagger_\sigma|\Gamma'\rangle\,\phi^\dagga_{\Gamma' m}\,
\langle m|d^\dagga_{a\sigma}|n\rangle \nonumber\\
&=& \fract{1}{\sqrt{n^0_{a\sigma}\left(1-n^0_{a\sigma}\right)}}\sum_{\Gamma\Gamma' n m}\, 
\phi^\dagger_{n\Gamma}\,\langle \Gamma|c^\dagger_\sigma|\Gamma'\rangle\,\phi^\dagga_{\Gamma' m}\,
\langle m|d^\dagga_{a\sigma}|n\rangle \label{Z-c-final},
\eea
\end{widetext} 
which are analogous to those proposed by Lechermann and coworkers~\cite{Georges}, as shown 
by B\"unemann and Gebhard.~\cite{quaquaraqua}  

\subsection{The variational energy}

The average value of the Hamiltonian \eqn{Ham} is the sum of two terms, the average of the hopping 
$\mathcal{H}_0$ plus that of the Kondo exchange $\mathcal{H}_J$. The latter is a purely local term, hence, 
provided Eqs.~\eqn{uno-app-bis} and \eqn{due-app-bis} are verfified and in infinite-coordination lattices , 
can be written through \eqn{av-O-final} as (we reintroduce the site label $\bR$) 
\be
\Av{\Psi}{\mathcal{H}_j} = J\,\sum_\bR\,\Tr\Bigg(\phi^\dagger(\bR)\,\mathbf{S}_{f\bR}\cdot\mathbf{S}_{c\bR}\,
\phi(\bR)\Bigg).\label{av-J}
\ee
A way to proceed is to use as variational parameters the matrices $\phi(\bR)$ and the $n^0_{a\bR\sigma}$'s, which are 
related to each other by the conditions Eqs.~\eqn{uno-app-bis} and \eqn{due-app-bis}. Then, the Slater 
determinant $|\Psi_0\rangle$ must be the one that minimizes the average value of the hopping, which is, 
through Eqs.~\eqn{Z-qp} and \eqn{Z-c-final},  
\begin{widetext}
\be
\Av{\Psi}{\mathcal{H}_0} = -t\sum_{<\bR\bRp>\sigma}\,\sum_{a,b=1}^2 \Bigg( 
Z_{ca\sigma}(\bR)\,Z^*_{cb\sigma}(\bRp)\,
\Av{\Psi_0}{d^\dagger_{a \bR \sigma}d^\dagga_{b \bRp \sigma}} + c.c.\Bigg),
\ee
\end{widetext}     
under the constraint that the single particle density matrix has eigenvalues $n^0_{a\bR\sigma}$.
One readily realizes that $|\Psi_0\rangle$ that fullfills such a property is actually the ground state of the following variational Hamiltonian 
\begin{widetext}
\bea
\mathcal{H}_* &=&  -t\sum_{<\bR\bRp>\sigma}\,\sum_{a,b=1}^2\,\Bigg( 
Z_{ca\sigma}(\bR)\,Z^*_{cb\sigma}(\bRp)\,
d^\dagger_{a \bR \sigma}d^\dagga_{b \bRp \sigma} + H.c.\Bigg) \nonumber\\
&& - \sum_\bR\,\sum_{a,b=1}^2\,\sum_\sigma\, \Bigg[\mu_{ab\sigma}(\bR)\,
\bigg(d^\dagger_{a\bR\sigma}d^\dagga_{b\bR\sigma} - \delta_{ab}\,n^0_{a\bR\sigma}\bigg) 
+ H.c.\Bigg],\label{H-variational}
\eea
\end{widetext}
where the Lagrange multipliers $\mu_{ab\sigma}(\bR)$ are those that maximize the ground state energy, so that 
\be
\Av{\Psi}{\mathcal{H}_0} = \Av{\Psi_0}{\mathcal{H}_*}.\label{E_*}
\ee
Once this task has been accomplished, one needs to minimize the variational energy per site 
\bea
&& E_{var} = \frac{1}{N}\Av{\Psi_0}{\mathcal{H}_*} \nonumber \\
&& ~~~~~ +  J\,\sum_\bR\,\Tr\Bigg(\phi^\dagger(\bR)\,\mathbf{S}_{f\bR}\cdot\mathbf{S}_{c\bR}\,
\phi(\bR)\Bigg),\label{E-var-app}
\eea
with respect to $\phi(\bR)$ and $n^0_{a\bR\sigma}$ that fulfill Eqs.~\eqn{uno-app-bis} and \eqn{due-app-bis}.

\subsection{Technical remarks}

In order to parametrize the variational matrix $\phi(\bR)$ (in what follows we assume a generic multiband Hamiltonian) 
one can introduce a local Hamiltonian $h_\bR$ that acts on all possible local electronic configuration, and define 
\be
\phi(\bR)^\dagger \phi(\bR) = \fract{\mathrm{e}^{-\beta h_\bR}}{\Omega_\bR},\label{def-hR}
\ee
where 
\[
\Omega_\bR = \Tr\left(\mathrm{e}^{-\beta h_\bR}\right),
\]
is the local partition function and $1/\beta$ a fictitious temperature. With this definition, the condition 
Eq.~\ref{due-app-bis} becomes 
\be
C_{ab}(\bR)\equiv \fract{1}{\Omega_\bR}\, \Tr\left(\mathrm{e}^{-\beta h_\bR}\,c^\dagger_{a\bR} c^\dagga_{b \bR} 
\right) = \langle \Psi_0| c^\dagger_{a\bR} c^\dagga_{b \bR} |\Psi_0\rangle,\label{C_ab}
\ee
where $a$ and $b$ label both spin and orbitals. Therefore, the zero-temperature average value of the single-particle 
density matrix on $|\Psi_0\rangle$ must coincide with the its thermal average with the local Hamiltonian $h_\bR$.
Given $h_\bR$, one calculates $C_{ab}(\bR)$, Eq.~\eqn{C_ab}, which can be diagonalized providing the definition of the 
natural basis:
\be  
\fract{1}{\Omega_\bR}\,\Tr\left(\mathrm{e}^{-\beta h_\bR}\,d^\dagger_{a\bR} d^\dagga_{b \bR} 
\right) = \delta_{ab}\,n_{a\bR}.
\ee
In terms of $h_\bR$
\be
\phi(\bR) = U^\dagga_\bR\, \fract{\mathrm{e}^{-\beta h_\bR/2}}{\sqrt{\Omega_\bR}},
\ee
with $U^\dagga_\bR$ a unitary matrix. The expressions of the $Z$ renormalization factors are then obtained through 
\bea
&&\fract{1}{\Omega_\bR}\, \Tr\Bigg(\mathrm{e}^{-\frac{\beta}{2} h_\bR}\, U^\dagger_\bR\, c^\dagger_{a\bR}\, 
U^\dagga_{\bR}\, \mathrm{e}^{-\frac{\beta}{2} h_\bR}\, d^\dagga_{b\bR}\Bigg)\nonumber\\
&& = Z_{ab}(\bR)\,\sqrt{n_{b\bR}\left(1-n_{b\bR}\right)}.\label{App-Z}
\eea
We found that it is more convenient to use as variational parameters instead of the matrix elements of $\phi(\bR)$ those of 
the local Hamiltonian $h_\bR$ and of the unitary matrix $U_\bR$. In the case of a paramagnetic wavefunction that does not 
break translationally symmetry, $h_\bR$ and $U_\bR$ are independent of $\bR$. On the contrary, for antiferromagnetic wavefunctions on a bipartite 
lattice, going from one sublattice to the other the role of spin $\up(\giu)$ is interchanged with that of spin $\giu(\up)$.

%\bibliographystyle{apsrev}
%\bibliography{biblio}

\begin{thebibliography}{43}
\bibitem[{not({\natexlab{b}})}]{nota-Ogata}
\bibinfo{note}{In a square lattice with nearest neighbor hopping $t$, half the
  bandwidth $D=4t$, so the values of $J/D$ in Fig.~\ref{phd-AF} must be
  multiplied by 4 to be compared with the values of $J/t$ in Fig.~2 of
  Ref.~\onlinecite{Ogata}}.
\bibitem[{sas()}]{sasso}
\bibinfo{note}{In particular, Eq.~(26) in Ref.~\onlinecite{quaquaraqua} is
  wrong. With our notations, this equation reads \[ \langle \Psi_0|\,
  d^\dagger_{a\alpha}d^\dagga_{b\beta}\, |n\rangle\langle m|\,|\Psi_0\rangle
  \stackrel{?}{=} \langle m|\,
  d^\dagger_{a\alpha}d^\dagga_{b\beta}\,|n\rangle\,
  \sqrt{P^0_n\,P^0_m},\label{wrong} \] which is not correct since the right
  result is instead \[ \langle \Psi_0|\, d^\dagger_{a\alpha}d^\dagga_{b\beta}\,
  |n\rangle\langle m|\,|\Psi_0\rangle = \langle m|\,
  d^\dagger_{a\alpha}d^\dagga_{b\beta}\,|n\rangle\, P^0_m.\label{true} \]}.
\end{thebibliography}

\end{document}